\documentclass[prl, preprint, nobibnotes,twocolums, showpacs]{revtex4}
\usepackage{color,graphicx}
\begin{document}

\title{$Graphitic$-BN Based Metal-Free Molecular Magnets From A First Principle Study}

\author{R. Wu, L. Liu, Y. P. Feng}
\email{phyfyp@nus.edu.sg}
\affiliation{Department of Physics, National University of Singapore, 2 Science Drive 3, Singapore 117542}

\begin{abstract}
We perform a first principle calculation on the electronic properties of carbon doped {\em graphitic} boron nitride ({\em graphitic}-BN). It was found that carbon substitution for either boron or nitrogen atom in {\em graphitic}-BN can induce spontaneous magnetization. Calculations based on density functional theory with the local spin density approximation on the electronic band structure revealed a spin polarized, dispersionless band near the Fermi energy. Spin density contours showed that the magnetization density originates from the carbon atom. The magnetization can be attributed to the carbon $2p$ electron. Charge density distribution shows that the carbon atom forms covalent bonds with its three nearest neighbourhood. The spontaneous magnetization survives the curvature effect in BN nanotubes, suggesting the possibility of molecular magnets made from $graphitic$-BN. Compared to other theoretical models of light-element or metal-free magnetic materials, the carbon-doped BN are more experimentally accessible and can be potentially useful.
\end{abstract}


\maketitle
  Tranditional magnetic metal based materials have a Curie temperature not over 1,400 {\em K}. Thus there has been a long standing search for novel high temperature magnets\cite{Palacio}. Potential magnetism from metal-free system was probably firstly proposed by Fujita {\em et al.} \cite{Fujita}. they suggested the possible magnetic structure in the nano-scaled zigzag graphite ribbon resulting from the effect of localized edge states. The weak ferromagnetism in all-carbon system were observed by Makarova \cite{Makarova} and Esquinazi {\em et al.} \cite{Esquinazi} respectively. These studies stimulated the interest to explore the possible mechanism of the unexpected magnetism in matal-free system. Lehtinen {\em et al.} \cite{Lehtinen} performed an {\em ab initio} local-spin-density approximation (LSDA) calculation to study the properties of a carbon adatom on a graphite sheet and found that this defect is magnetic with a moment up to 0.5 $\mu_B$. This was important in understanding the magnetism experimentally observed by Esquinazi \cite{Esquinazi} in the graphite system. More recently, Ma {\em et al.}\cite{Ma} studied the magnetic properties of vacuncies in graphite and carbon nanotubes. They found that the vacancy was spin-polarized with a magnetic moment of 1.0 $\mu_B$ in the graphite sheets and also introduce magnetism in the carbon nanotubes, depending on the chirality of the nanotubes and the structural configuration with respective to the tube axis.

The unexpected magnetism in metal-free system also ignited ambitions to search for metal-free magnetic materials in view of their potential applications as high temperature magnetis since metal magnets lose their ferromagnetism at high temperatures. Based on the results of Fjuita {em et al.} \cite{Fujita}, Kusakabe and Maruyama \cite{Kusakabe} further predicted that the nanoscaled zigzag-edged graphite ribbon could also have finite magnetization when each carbon at one edge is hydrogenated by two hydrogen atoms while each carbon atom at the other edge is hydrogenated with a single hydrogen atom. Another recently proposed model by Choi and coworkers\cite{Choi} is the heterostructured C-BN nanotubes. They calculated the electronic structure of the (9,0) C$_1$(BN)$_1$ and C$_2$(BN)$_2$ nanotubes using density functional theory and found the occurrence of magnetism at the zigzag boundary connecting carbon and boron nitride segments of tubes. However, in view of their structural configurations, experimental synthesis was a great challenge. For Kusakabe's model, fully hydrogenating one edge while halfly hydrogenating the other side uniformly could not be done in the same experimental condition. what's more, in the half hydrogenation, it was hard to prevent two hydrogen atoms hydrogenating a single carbon atom which would destroy the magnetic ordering.  As for the model of Choi \cite{Choi}, it required a BN nanotube segment to replace carbon compartment along the axis direction to form a C-BN nanotube superlattice which is very hard in an enviroment where the three elements coexist since they could form covalent compound with each other. 

  However, all these theoretical calculation had throught some hints about the nature of the magnetsim in the carbon-based system. That is, the magnetism may result from undercoordinated carbon atoms \cite{Lehtinen, Fujita} or unpaired carbon $2p$ electrons \cite{Choi,Kusakabe}. In this report we explore the electronic properties of carbon-doped $graphitic$-BN. The reason lies in the follwing reasons. First, $graphtic$-BN has a melting point over 1,700 {\em K}, which make it suitable for high temperature applications; Second, $graphitic$-BN can form different nano-scaled structures such as nanotubes\cite{Chopra}, fullerenes\cite{fullerene} and fullerites \cite{fullerites}. The BN-based magnetic nano-structures can serve as molecular magnets; Third, most important of all, in the hexgonal network, a carbon atom may form three bonds with either boron or nitrigen atoms, leaving one electron unpaired (the interactions between layers are only weak Van der Waals, so the effect can be ignored) , making it a possibility of spontaneous magnetization in such a carbon containing boron nitride system. 

  Our calculation is base on the spin-polarized density functional theory (DFT). The generalized gradient approximation (GGA) of Perdew and Wang \cite{Perdew} is adaptted for the exchange-correlation potential. The projector augmented wave (PAW) potentials have been used to represent the electron-ion interactions. All 2$s$ and 2$p$ electrons are considered as valence electrons. A kinetic energy cutoff of 520 eV is used to ensure a convergence better than 1 meV for total energies. A (3$\times$3 $\times $1) $graphitic$-BN sheet seperated by a 12 \AA\ vacuum under periodical bourdary conditions (PBC)is used for calculations. A Gamma-centered (4$\times $4$\times $1) K-mesh was used to sample the irreversible Brillouin Zone (IBZ). The conjugate-gradient algorithm \cite{cga} is applied to relax the ions into their intantaneous ground states. The structural parameters have been adjusted so that the external pressure is less than 1 kBar along each lattice direction. We first calculate the structural and electronic properites of $graphitic$-BN and compare them to previous theoretical calculations and experimental studies before the calculations on the carbon doped supercell. All the calculation was done using the plane wave basis VASP package \cite{vasp}. 

  We first calculated the lattice parameters and band structure of $graphtic$-BN sheet.  Our calculation reached a lattice constant of 2.52 \AA\ and an indirect band gap of 4.61 eV. This reproduced previous calculation by Xu \cite{Xu}. The validified the application of the projector augmented wave (PAW) potentials. The formation energy $E_{form}$ is defined as   

  $E_{form}$ = $E_{tot}$(doped)-$E_{tot}$(perfect)-($E_A$(isolated)-$E_B$(isolated)

where $E_A$(isolated) and $E_B$(isolated)is the energy of doping atom and the substituted atom at isolated states respectively. The calculated $E_{form}$ is 1.95 eV for carbon substitution upon boron and 3.46 eV for carbon substition upon nitrigen. Compared to the energy required to dope a silicon atom substitution upon a carbon atom in C$_{60}$ \cite{c59si} which is 5.77 eV and the experimental observation of C$_{59}$Si$_1$ \cite{c59siexp}, the energy required for a carbon atom to substitute boron or nitrogen atom in $graphtic$-BN is still within experimental access.

  Fig.~1 (a) is the band structure for the $graphitic$-BN with a boron atom substituted by carbon atom. We can see that the carbon substitution caused a lift up of the Fermi level of the system and the occurance of a flat band near Fermi level. This is because the carbon atom acts as a $n$-type dopant since a carbon atom consists 4 valence electrons. All bands with energies lower thatn 4 eV are fully occupied and thus have no contribution to the spin-polarization. However, the flat band near Fermi level is split into two branches. The spin-up branch is occupied and the spin-down branch is left empty, leading to a spontaneous polarization in the doped system with a net magnetic moment of around 1 $\mu_B$. The same effect also survive when a nitrogen atom is substituted by carbon atom (Fig.~1 (b)). The only difference is that the Fermi level is push down ( in this case, the carbon atom acts as $p$-type dopant because it have less one electron than nitrogen atom) and the occupied branch merges into those fully occupied bands. 

  To find out the origin of the magnetism, we plot the magnetization density (spin-up minus spin-down density) contours through the sheet for both system in Fig.~2. As we can see that the magnetization density originates from where the carbon atom resides. No magnetization density is found at other atomic sites. This suggests that the spin-polarization results from the unpaired electron in the carbon atom. In Fig.~3, we plot the spin density distribution contour perpendicular to the sheet plane. In both system the spin density showed a $p$-orbital character. These results are reasonable as in the hexagonal network, a single carbon atom has three nearest neighbourhood to form bonds with, leaving one electron unpaired. It is obvious that in both case, the magnetization is from one of the carbon 2$p$ electrons. 

  One question in concern is that whether the carbon dopant can form bonds with its three nearest neighbourhood and maintain the fine structure. There has been successful experimental synthesis of the graphitic BC3 honeycomb epitaxial sheet\cite{BC3}. We plot the charge density distribution contour through the carbon doped BN sheet in both cases shown as Fig.~4 (a) and (b). In both systems we can see that carbon dopant forms bonds with nearest neighbourhood. When carbon is surrounded by nitrogen atoms (a), the electron clouds are localized around the connecting lines between carbon and nitrogen atoms. Due to the difference of electronegativity, carbon looses part of it valence electrons. In the case of substitution for nitrogen (b), carbon atom attracts electrons from boron atoms. In both case, the nature of the bonds is covalent and no structural distortion has been observed opun geometry optimization. We also find that the spontaneous spin-polarization survive the curvature effect in the BN nanotubes \cite{arxiv}. Thus the various nanostructures originating from $graphitic$-BN promise to act as molecular magnets when doped with carbon. Contrast to the models proposed by Kusakabe \cite {Kusakabe} and Choi \cite {Choi} in which the spin polarization only occurs for certain particular structural configurations, the spin polarization in $graphitic$-BN is induced by substitution with arbitrary atomic configuration, indicating easy accessibility of experimental synthesis.

Boron nitride has important advantages over carbon nanostructures. It is far more resistant to oxidation than carbon and therefore more suitable for high-temperature applications in which carbon nanostructures would burn.
Unlike carbon nanostructures, BN based nanostructures are insulators, with predictable electronic properties. In addition,
magnetic nanostructures are of scientifically interesting and technologically important, with many present and future applications in permanent magnetism, magnetic recording and spintronics. The carbon substituted BN nanostructure, with conduction electrons that are 100\% spin polarized due to the gap at the Fermi level in one spin channel, and a finite density of states for the other spin channel, can be an ideal half metallic material and can be useful for spintronic applications, such as tunneling magnetoresistance and giant magnetoresistance elements.

In conclusion, we have performed first principle projector augmented wave (PAW) potential calculation with  plane wave basis sets to study the effects of carbon substitution on $graphitic$-BN. Our results show that single carbon atom substitution for any atom results in a polarized flat band and a sharp peak in the majority density of state below the Fermi level. 
The spontaneous spin-polarization is independent of site of substitution. Compared to other metal-free magnets previously proposed, the carbon induced magnetization in $graphitic$-BN is experimentally accessible and the system can be potentially very useful.

\newpage
\begin{figure}[hbt]
\includegraphics{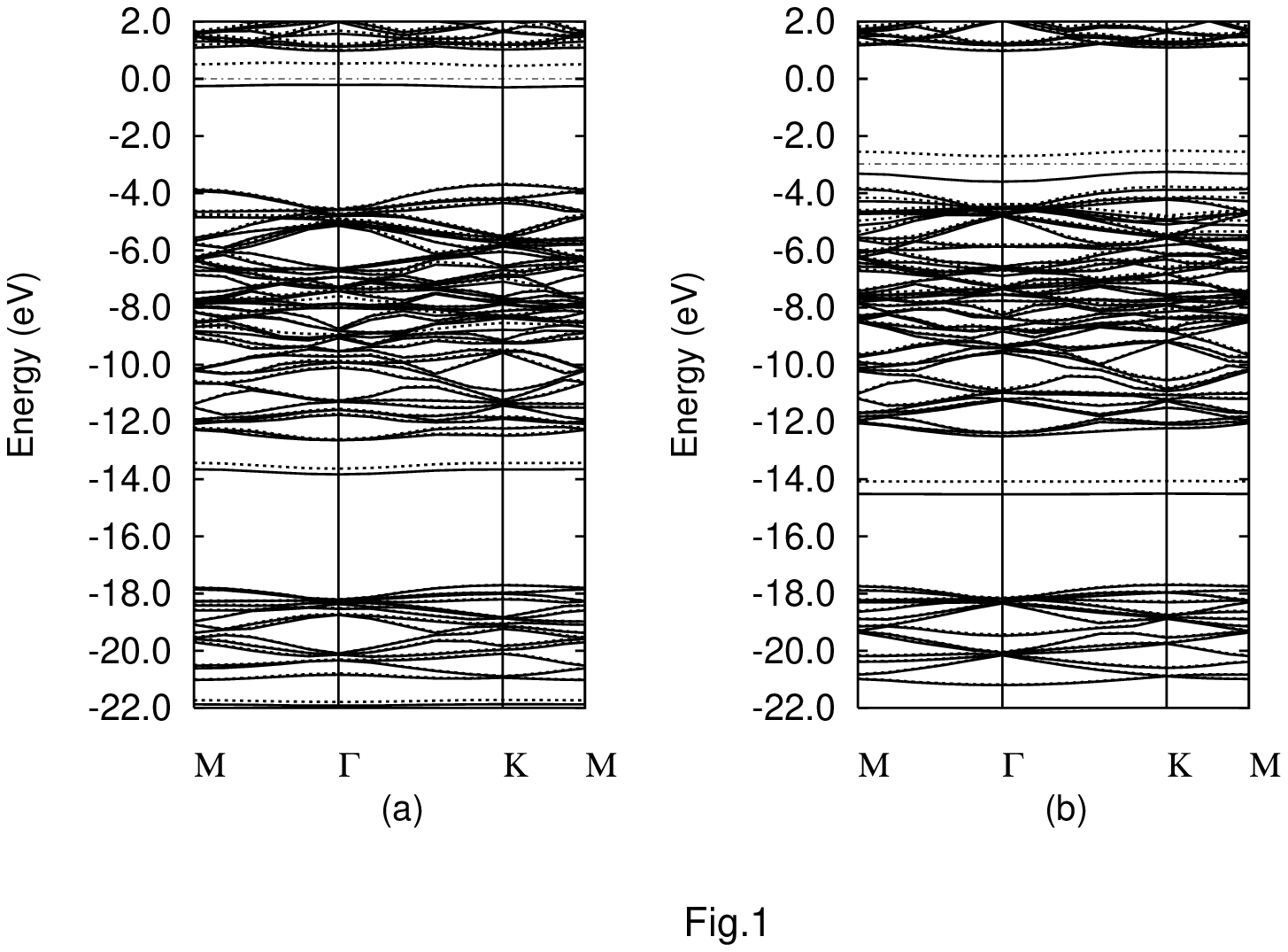}
\caption{The calculated band structures of the $graphitic$-BN with a boron atom (a) and a nitrogen atom (b) substituted by a carbon atom. The solid lines represent the spin-up branches while the dot lines represent the spin-down branches. Fermi level is denoted by the light dot line.}
\end{figure}

\begin{figure}[hbt]
\includegraphics{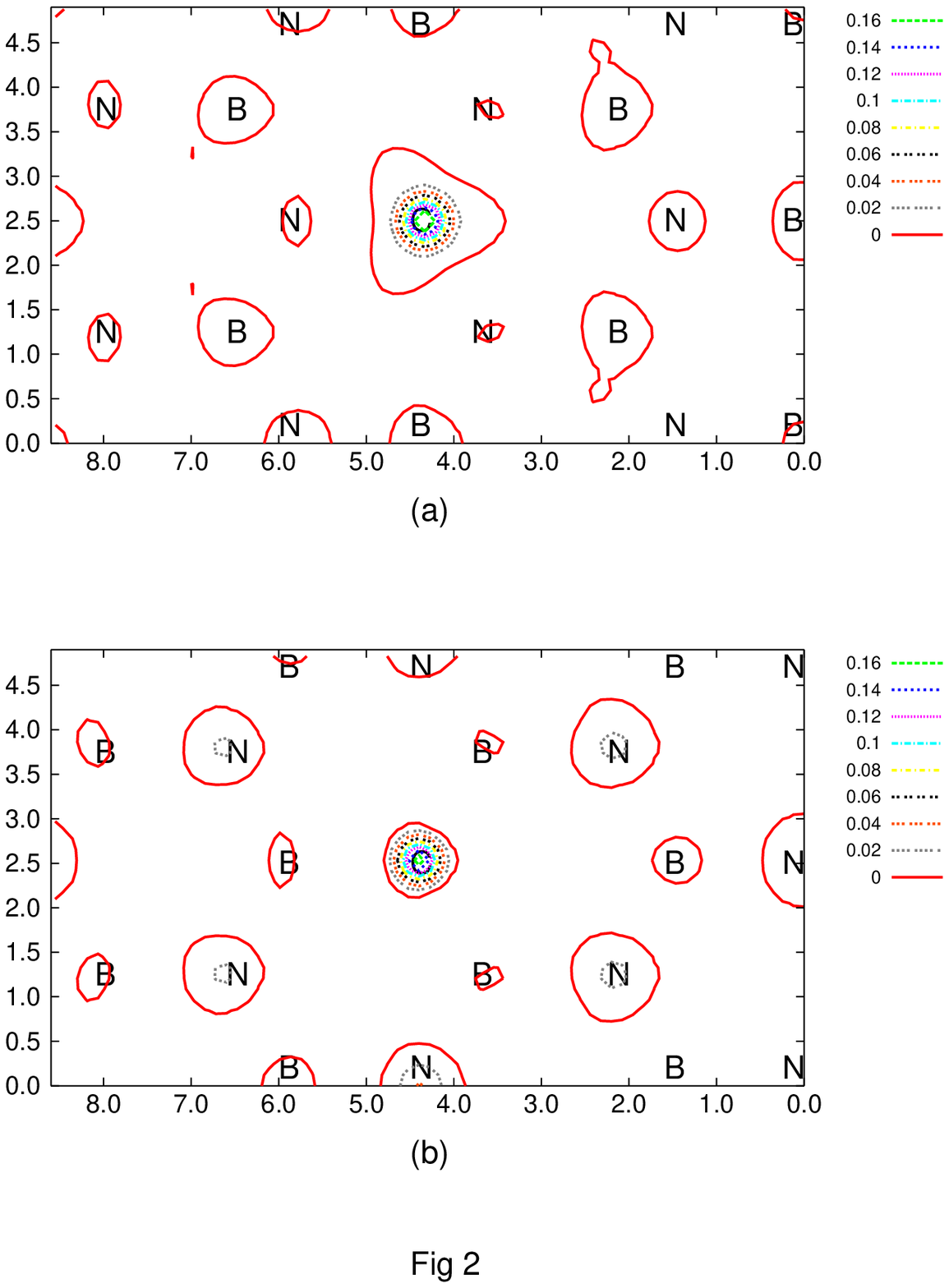}
\caption{The magnetization density (spin-up minus spin-down) distribution of the $graphitic$-BN with a boron atom (a) and a nitrogen atom (b) substituted by a carbon atom through the sheet plane. The unit: spins/\AA $^3$} 
\end{figure}

\begin{figure} [hbt]
\includegraphics{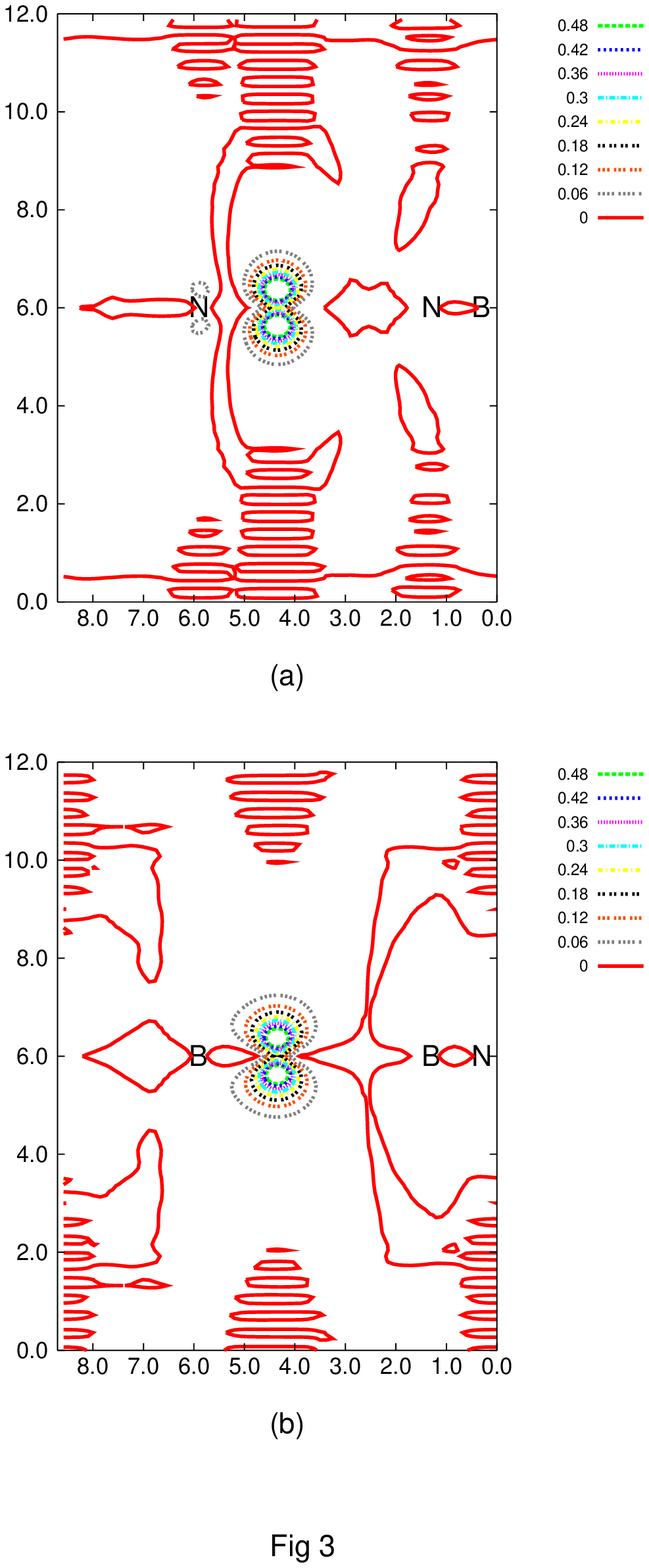}
\caption{The magnetization density (spin-up minus spin-down) distribution of the $garphitic$-BN with a boron atom (a) and a nitrogen atom (b) substituted by a carbon atom perpendicular to the sheet plane. The unit: spins/\AA $^3$} 
\end{figure}

\begin{figure}[hbt]

\caption{The charge distribution of the $graphitic$-BN with a boron atom (a), and a nitrogen atom (b) substituted by a carbon atom. The unit: electrons/\AA $^3$ }
\includegraphics{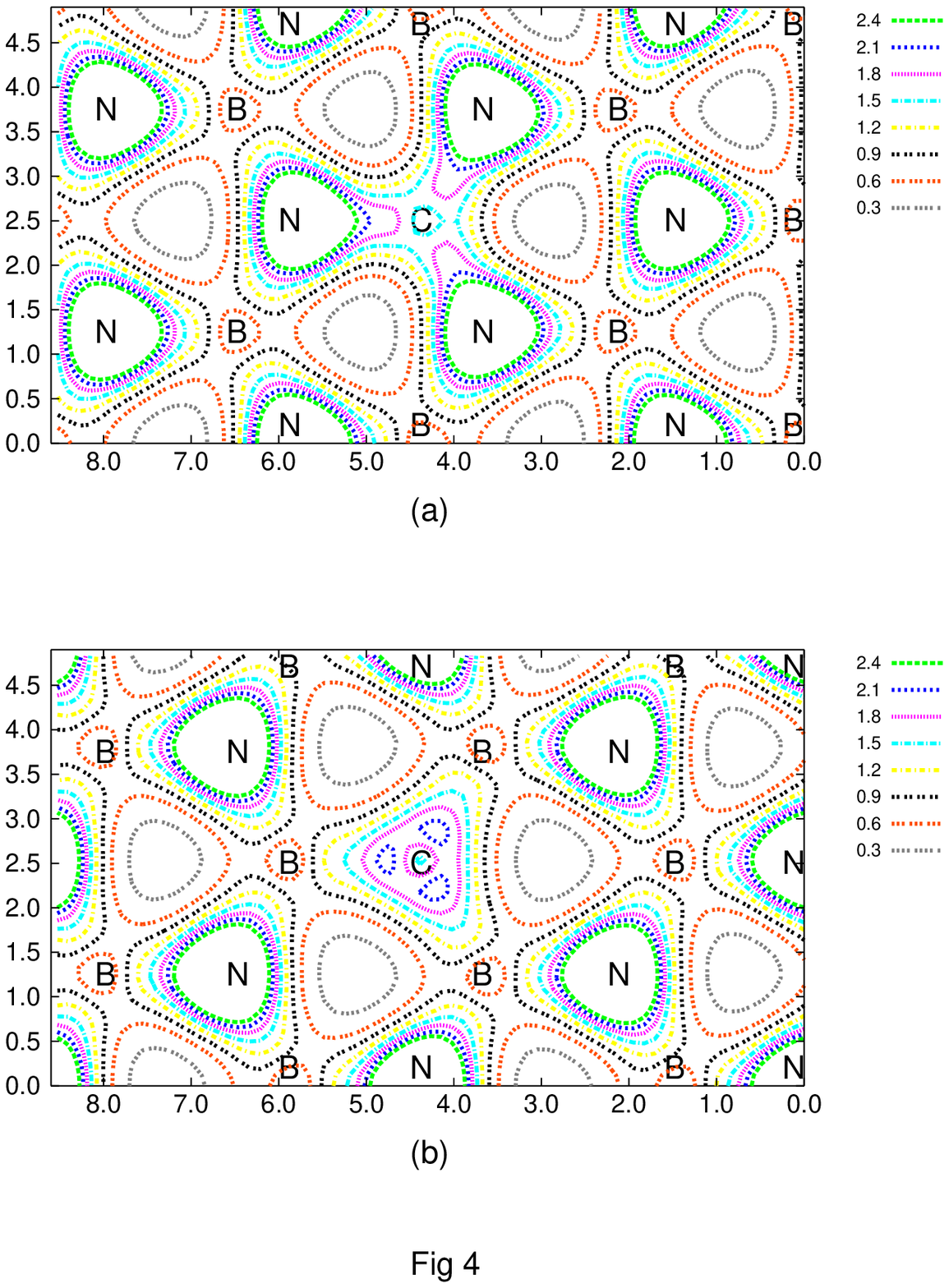}
\end{figure}


\newpage
\begin{thebibliography}{}

\bibitem{Palacio}
F. Palacio, Nature (London) {\bf 413},690 (2001), and references therein

\bibitem{Fujita}
M. Fujita, K. Wakabayashi, K. Nakata and K. Kusakabe, J. Phys. Soc. Jap {\bf 65}, 1920 (1996).
 
\bibitem{Makarova}
T. L. Makarova, B. Sundqvist, R. H\"ohne, P. Esquinazi, Y. Kopelevich, P. Scharff, V. A. Davydov, L. S. Kashevarova, and A. V. Rakhmanina, Nature, {\bf 413}, 716 (2001).

\bibitem{Esquinazi}
P. Esquinazi, A. Setzer and R. Hohne, C. Semmelhack, Y. Kopelevich, D. Spemann, T.  Butz, T, B. Kohlstrunk and M. Losche, Phys. Rev. B {\bf 66}, 024429 (2002).

\bibitem{Lehtinen}
P. O. Lehtinen, A. S. Foster, A. Ayuela, A. Krasheninnikov, K. Nordlund and R. M. Nieminen, Phys. Rev. Lett {\bf 91}, 017202 (2004)£»Lehtinen, P.O., Foster, A.S., Ayuela, A., Vehvil\"ainen, T.T. and Nieminen, R.M£¬ phys. Rev. B {\bf 69}, 155422 (2004).

\bibitem{Ma}
Yuchen Ma, P. O. Lehtinen, A. S. Foster and R. M. Nieminen, New J. Phys {\bf 6}, 68 (2004).

\bibitem{Kusakabe}
K. Kusakabe and M. Maruyama, Phys. Rev. B {\bf 67}, 092406 (2003).

\bibitem{Choi}
J. Choi, Y. H. Kim, K. J. Chang and D. Tomenek, Phys. Rev. B {\bf 67}, 125421 (2003).

\bibitem{Chopra}
N. G. Chopra, R. J. Luyken, K. Cherrey, V. H. Crespi, M. L. Cohen, S. G. Louie and A. Zettl, Science {\bf 269}, 966 (1995) 

\bibitem{fullerene}
D. Golberg, Y. Bando, O. Stephan and K. Kurashima, Appl. Phys. Lett, {\bf 73}, 2441 (1998)

\bibitem{fullerites}

Pokropivny, V. V. A; Skorokhod, V.V; Oleinik, G.S; Kurdyumov, A.V; Bartnitskaya, T.S; Pokropivny, A.V; Sisonyuk, A.G and Sheichenko, D.M, J. Solid. state. Chem, {\bf 154}, 214 (2000) 

\bibitem{Perdew}
J. P. Perdew, J. A. Chavery, S. H. Vosko, K. A. Jackson, M. R. Pederson, D. J. Singh and C. Foilhais, Phys. Rev. B {\bf 46}, 6671 (1992)
\bibitem{cga}
W. H. Press, B. P. Flannery, S. A. Teukolsky and W. T. Vetterling, New Numerical 
Recipes (Cambridge University Press, New York, 1986).


\bibitem{vasp}
G. Kress and J. Furthm\"uller, Comput. Mater. Sci. {\bf 6}, 15 (1996)

G. Kress and J. Furthm\"uller, Phys. Rev. B {\bf 54}, 11169 (1996)

\bibitem{Xu}
Yong Nian Xu and W. Y. Ching, Phys. Rev. B {\bf 44}, 7787 (1991)

\bibitem{c59si}
Chu-chun Fu, Mariana Weissmann, Maider Machado, Pablo Oderj{\"o}n,  Phys. Rev. B {\bf 63}, 85411 (2001)

\bibitem{c59siexp}
C. Ray, M. Pellarin, J. L. Lerm{\'e}, L. J. Vialle and M. Broyer, Phys. Rev. Lett. {\bf 80}, 5365 (1998)

\bibitem{BC3}
H. Yanagisawa, T. Tanaka, Y. Ishida, M. Matsue, E. Rokuta, S. Otani and C. Oshima, Phys. Rev. Lett, {\bf 93}, 177003 (2004)
 
\bibitem{arxiv}
R. Wu, L. Liu, G. Peng and Y. P. Feng, accepted by Appl. Phys. Lett
 
  
\end{thebibliography}
\end{document}